# Longitudinal Magnetic Dynamics and Dimensional Crossover in the Quasi-One-Dimensional, Spin-1/2, Heisenberg Antiferromagnet KCuF$_3$

B. Lake[1,2], D.A.Tennant[3,4*] and S.E. Nagler[1]

[1] Center for Neutron Scattering, Oak Ridge National Laboratory, Oak Ridge, Tennessee 37831-6393, USA.

[2] Clarendon Laboratory, Parks Road, Oxford OX1 3PU, United Kingdom.

[3] ISIS Facility, Rutherford Appleton Laboratory, Chilton, Didcot, Oxon OX11 0QX, United Kingdom.

[4] School of Physics & Astronomy, North Haugh, St Andrews, Fife KY16 9SS, U.K.

**Abstract**

The spin dynamics of coupled spin-1/2, antiferromagnetic Heisenberg chains is predicted to exhibit a novel longitudinal mode at low energies and temperatures below the Néel temperature. This mode is a dimensional crossover effect and reveals the presence of a limited amount of long-range antiferromagnetic order co-existing with quantum fluctuations. In this paper the existence of such a mode is confirmed in the model material KCuF$_3$ using polarized and unpolarized inelastic neutron scattering and the longitudinal polarization of the mode is definitively established. The lineshape is broadened suggesting a reduced lifetime due to decay into spin-waves. In addition the data shows evidence of continuum scattering with a lower edge greater than the longitudinal mode energy. A detailed comparison is made with theoretical predictions and experimental work on other model materials.

## 1. Introduction

The physical behaviour of the spin-1/2 ($S$=1/2) Heisenberg antiferromagnetic chain (HAFC), defined by the simple Hamiltonian

$$H_{1D} = J\sum_i \vec{S}_i \cdot \vec{S}_{i+1} \tag{1}$$

where $i$ is the site index along the chain and $J$ is the antiferromagnetic exchange constant, is highly non-trivial. It displays a spin singlet ground state with algebraically decaying correlations, and its excitations are free spinons i.e. quantized π-solitons carrying spin quantum numbers of 1/2, rather than the $S$=1 carried by conventional spin-waves. Spinons are semions obeying half-fractional statistics, and are restricted to creation in pairs [1,2]. The spinon picture has been confirmed in some detail by measurements of the triplet excitation continuum in KCuF$_3$ [3,4,5], and other materials [6,7,8,9] and the dynamics of an isolated $S$=1/2 HAFC are relatively well understood theoretically and experimentally. However measurements on quasi-one-dimensional (quasi-1D) materials at low energy and temperature scales where effects of interchain coupling become significant have shown surprising and non-trivial crossover effects. In particular in our recent studies of KCuF$_3$ a damped mode is observed in a region of wavevector and energy space where the three-dimensional (3D) character of the excitations is important [10,11,12]. This mode was interpreted as a longitudinal mode.

A long-lived longitudinal magnetic excitation was predicted in some weakly-coupled cluster compounds e.g. the $S$=1/2 dimer [13] and tetramer [14] and has been measured using Raman scattering in the tetramer system

---

Cu$_2$Te$_2$O$_5$Br$_2$ [14]. In quasi-1D antiferromagnets the stability of a longitudinal mode was investigated theoretically for integer-spin chains using quantum field-theory by Affleck and Wellman [15] in response to anomalous experimental results on CsNiCl$_3$ [16,17]. In both these cases an anisotropy or magnetic field is able to mix states and split a pre-existing triply degenerate magnon into transverse and longitudinal components. However for half-odd-integer spins a topological term in the field theory substantially alters the physics so that the basic excitations are fractional particles known as spinons which are observed as a multi-spinon continuum rather than a well-defined magnon mode. Nevertheless, inspired by experimental studies of KCuF$_3$ [3,4,5] Schulz [18] demonstrated, using field theoretical techniques, that a stable longitudinal mode may arise for the case of spin-1/2 chains coupled weakly together into a tetragonal lattice (the minimal model for KCuF$_3$).

The $S=1/2$ HAFC Hamiltonian (equation (1)), extended to include coupling between neighboring chains becomes,

$$H_{Q1D} = J\sum_{i,r} \vec{S}_{i,r} \cdot \vec{S}_{i+1,r} + J_\perp \sum_{i,r} \left( \vec{S}_{i,r} \cdot \vec{S}_{i,r\pm a} + \vec{S}_{i,r} \cdot \vec{S}_{i,r\pm b} \right) \qquad (2)$$

where $J_\perp$ is the interchain exchange constant and $r$ is a lattice vector in the $a$-$b$ plane (perpendicular to the chains that run along the $c$ direction) which is used to label the magnetic sites. $J_\perp$ couples nearest neighbour chains only and for a material with tetragonal symmetry the strength of $J_\perp$ is the same in the $a$ and $b$ directions. Each chain site then has two neighbours along $c$, and four within the $a$-$b$ plane. In the Hamiltonian of equation (2) the couplings are particularly simple, being both unfrustrated and three-dimensional. Combined with the mathematical tractability of the spin-1/2 HAFC, equation (1), which has a quantum critical ground state and displays scaling behavior, Hamiltonian (2) is an excellent model to explore dimensional crossover effects on strongly fluctuating quantum states. Numerous theoretical works, based on systematic treatments of the interchain coupling in equation (2), have subsequently been published, furnishing detailed predictions for static and dynamical properties including for the longitudinal mode, and these have been compared to experimental data from KCuF$_3$, and other model magnetic materials.

The most comprehensive experimental and theoretical work to date, has been directed toward KCuF$_3$. Using unpolarized neutron scattering the dynamical response for all energy, wavevector and temperature scales has been sampled [10]. A damped mode attributed to the longitudinal mode was discovered in the 3D magnetically ordered phase with a wavevector and frequency dependence closely following predictions [11]. In this paper we provide additional details of the measurements reported in reference [11]. More importantly, we report a polarized neutron scattering study that definitively demonstrates the longitudinal polarization of the excitation reported there, and establishes the spin-polarizations of other features in the spectrum. This has been achieved by combining magnetic fields strong enough to pole the magnetic order of KCuF$_3$ into a single spin-flop domain, and exploiting inelastic neutron scattering measurements with vertical polarization analysis.

The paper is organized as follows: In section II, we summarize the magnetic properties of KCuF$_3$. Section III presents the primary theoretical predictions. The experimental method is explained in Section IV, and the results obtained from measurement and analysis presented in Section V. The wider significance of these results and their implications for other materials, namely BaCu$_2$Si$_2$O$_7$ where longitudinal excitations have also been investigated, are discussed in Section VI, where the conclusions are also given.

## II Magnetic properties of KCuF$_3$

The magnetic properties of KCuF$_3$ are near-ideal, and this was the first material where detailed experiments of higher-dimensional effects on spin-1/2 chains were conducted [19,11]. It has a tetragonal crystal structure (*Pnma*) that arises from a cooperative Jahn-Teller effect, with lattice parameters $a=b=4.126$Å and $c=3.914$Å (at $T=10$ K). The magnetic $d_{x^2-y^2}$ hole orbitals are ordered in an antiferrodistortive arrangement [5] that results in excellent quasi-1D properties and each Cu$^{2+}$ ion carries a near-isotropic spin ($S=1/2$) due to an almost complete quenching of the orbital angular momentum by the crystal field. The Cu $d$ orbitals are ordered, and there has been considerable work investigating this aspect of the structure, see e.g. [20]. A large orbital overlap through the fluorine $p$-orbital results in strong antiferromagnetic superexchange interactions ($J=34$ meV) coupling nearest neighbor magnetic sites in the ***c*** direction. This exchange is essentially an isotropic Heisenberg interaction with only a small (~0.2%) x-y anisotropy. As expected from Kanamori-Goodenough rules a much weaker ferromagnetic exchange interaction through an unfilled $d$-orbital ($J_\perp=-1.6$ meV) acts between sites in the ***a*** and ***b*** directions, coupling the Heisenberg chains together [21,22]. The interchain interactions induce long-range magnetic order below a Néel temperature of $T_N=39$ K - some 1/10$^{th}$ of the intrachain exchange strength $J/k$. The spin moments in the ordered phase are confined to the ***a-b*** plane by the x-y anisotropy with antiferromagnetic alignment along the chains (***c*** direction) and ferromagnetic alignment between the chains. Electron paramagnetic resonance and magnetic susceptibility measurements reveal a Dzyaloshinsky-Moriya interaction that destroys the in-plane degeneracy giving rise to an eight-sublattice magnetization structure where the spins are slightly canted away from the [1,1,0] and equivalent directions [23,24,25]. The ordered moment per spin $g_s\mu_B |\langle S[1,1,0]\rangle|$ obtained from neutron diffraction is 0.54$\mu_B$ [21] for temperatures $T\ll T_N$, indicating ~50% reduction of the magnetic order from its saturation value by zero-point quantum fluctuations.

The dynamics of KCuF$_3$ have been studied extensively: Above $T_N$, one-dimensional effects are found to dominate, and the spin dynamics measured by neutron scattering are in good agreement with that of free spinons [3,4,5], as approximated by the Müller ansatz [26]. Figure 1(a) illustrates the Müller ansatz at low energies around the antiferromagnetic zone centre, located at (0,0,1.5) in KCuF$_3$; scattering is expected within a V-shaped region centered at (0,0,1.5). This scattering is rotationally invariant i.e. the fluctuations take place equally in all directions. In the three-dimensionally ordered phase below $T_N$ the spin correlations at energies above 27 meV are not appreciably affected by the ordering. However, below this energy well-defined spin-wave modes exist with additional scattering lying between them for energies >12 meV. For wavevectors in the (0,0,$L$) direction conventional spin-wave theory with the inclusion of two-magnon terms could qualitatively explain the observations at low energies [19] but not the continuum at higher energies. Figure 1(b) shows the scattering calculated from spin-wave theory near (0,0,1.5) [19]. The spin-wave branches are well-defined transverse modes i.e. they involve oscillations of the magnetism in a direction perpendicular to the ordered spin moment, whereas the two-magnon signal is longitudinal (oscillations parallel to the direction of the spin ordering) and forms a broad continuum with a maximum at 23.5 meV and a full width at half maximum of 24.0 meV. Detailed measurements of low-energy fluctuations in KCuF$_3$ [11] and BaCu$_2$Si$_2$O$_7$ [27] have highlighted the need to include nonlinear effects in calculations of the dynamics.

## III Theoretical predictions

Systematic expansions in the interchain coupling in equation (2) have successfully described the static ordered moment, Néel transition temperature, and spin-waves measured at low temperatures within a unified theory, see e.g. [28]. Calculating the longitudinal dynamics though, presents the strictest test for theory. Recently, the $T=0$ K dynamics of the coupled $S=1/2$ Heisenberg antiferromagnet chain have been approached theoretically by considering the solution of the isolated chain in the continuum limit, treating the interchain interactions as a staggered field and applying the random phase approximation (RPA) [18,29]. The theory predicts a doubly degenerate, well-defined, transverse spin-wave mode with dispersion and intensity around the 1D antiferromagnetic zone centre given by

$$E_T = \sqrt{\frac{\pi^2 J^2}{4}(\pi n - 2\pi L)^2 + M^2\left(1 - \frac{\cos(2\pi H) + \cos(2\pi K)}{2}\right)} \quad (3)$$

and

$$\mathrm{Im}\,\chi_T = \frac{\pi}{2|J_\perp|}\delta\left(\frac{E^2}{M^2} - \frac{E_T^2}{M^2}\right) \quad (4)$$

respectively, where the wavevector transfer is $Q = (H, K, L)$, $M$ is the zone boundary energy for $L$ at the antiferromagnetic zone centre (e.g. (0.5, 0, 0.5)) which is related to $J_\perp$ and has the value 11 meV in KCuF$_3$, and $n$ is an odd integer such that $|n - 2L| \leq 0.5$. In addition the theory also predicts a well-defined longitudinal mode with energy and intensity,

$$E_L = \sqrt{\frac{\pi^2 J^2}{4}(\pi n - 2\pi L)^2 + M^2\left(3 - \frac{\gamma(\cos(2\pi H) + \cos(2\pi K))}{2}\right)} \quad (5)$$

and

$$\mathrm{Im}\,\chi_L = \frac{\pi\gamma}{4|J_\perp|}\delta\left(\frac{E^2}{M^2} - \frac{E_L^2}{M^2}\right) \quad (6)$$

where the constant $\gamma \approx 0.4913$. This mode has an energy gap at the antiferromagnetic zone center with gap size proportional to the interchain exchange constant $J_\perp$, and its intensity is a factor of four smaller than that of the transverse mode (for a given energy). The longitudinal mode contributes to the dynamics only when the ordered moment is suppressed by zero-point fluctuations and is therefore a quantum effect, its intensity drops away as the suppression decreases. Figure 1(c) shows the theoretical predictions for KCuF$_3$: The longitudinal mode lies between the dispersion branches of the transverse modes and has a gap of 17.4 meV. In addition Essler *et al.* predict continuum scattering starting at $2M=22$ meV and extending upwards in energy [29]. A broadened mode lying close to the theoretically predicted energy of the longitudinal mode was observed in the unpolarized neutron scattering measurements reported earlier [11].

**IV Experimental method**

Polarized and unpolarized inelastic neutron scattering measurements were carried out at the HB1 and HB3 triple axis spectrometers at the High Flux Isotope Reactor (HFIR) in Oak Ridge National Laboratory, U.S.A. The same high quality single crystal of KCuF$_3$ with a mosaic of 10´, mass of 6.86g and volume of 1cm³, that was employed in previous neutron measurements [3,4,5,19] was used.

*Unpolarized measurements*

Unpolarized inelastic neutron scattering measurements were made with a fixed final energy $E_F$=13.5 meV using the (0, 0, 2) reflection from a flat Pyrolytic Graphite (PG) analyser, and vertically focused PG monochromator. Higher order beam contamination was removed by a PG filter after the sample. The sample was mounted with its *a* and *c* lattice vectors horizontal in the scattering plane, and cooled to a base temperature of 2 K in a variable flow cryostat with temperature stability better than ±0.1 K. The steep spin-wave dispersion (200 meVÅ) along the [0,0,*L*] direction for KCuF$_3$ presents particular difficulties and a narrow wavevector resolution is required to resolve longitudinal signal between spin-wave branches. The 'resolution ellipsoid' must also have a 'vertical' orientation, so that when convolved with the spin-wave dispersion the two branches (with positive and negative slopes) have similar widths i.e. defocusing effects are small and the signal is symmetric about the antiferromagnetic zone centre. Measurements were conducted largely around (0, 0, 1.5) as it was found to be the best compromise between sharp, symmetric scans, and maximising the longitudinal magnetic intensity while limiting phonon contamination. A series of Soller collimators (values 48´-40´-40´-240´) between source and detector were selected to optimise conditions resulting in resolution limited spin-waves with full-width-half-maximum (FWHM) in wavevector of 0.057Å$^{-1}$ (0.035 r.l.u.), and an energy FWHM of 1.3 meV (for a dispersionless mode) at the predicted longitudinal mode energy (16 meV).

*Polarized measurements*

Distinguishing explicitly the transverse and longitudinal excitation components of a magnetic response using unpolarized neutron scattering is difficult. This however can be achieved with polarized neutron techniques. Using a vertical (up) polarizing monochromator and analyzer the polarization history of the detected neutrons can be determined. The scattering of up-spin neutrons scattering into up-spin final states is referred to as non-spin-flip (NSF) scattering while scattering into down-spin final states is designated spin-flip (SF) scattering. The addition of a flipping mechanism in the scattered beam allows for the independent detection of SF or NSF scattering. From this the vertical quantum number of the created excitation is also determined via conservation of spin moment. The main disadvantages of this technique are the large reduction in neutron intensity (compared to unpolarized neutron scattering) and the requirement of a mono-domain magnetic sample to achieve an unambiguous result. Because of the difficulty of such a study, this is the first polarization analysis of the spin dynamics in KCuF$_3$ to be carried out until now.

From Moon *et al* [30] the following rules for polarized neutron scattering can be abstracted: 1) Components of magnetic order or magnetic oscillations which are parallel to the direction of neutron polarization are seen in the NSF channel, 2) Components perpendicular to the direction of the neutron polarization are seen in the SF channel. 3) Phonons and structural scattering are observed only in the NSF channel. And finally, 4) those magnetic components which are parallel to the wavevector transfer are not observed in any channel. These rules can be combined to determine the polarizations of various magnetic excitations within a material, however to obtain an unambiguous result it is necessary to investigate a magnetically mono-domain sample so as to be able to define unique ordering and oscillation axes.

In zero-field, magnetic order in KCuF$_3$ is characterised by a structure where spin moments are slightly canted away from the [1,1,0] direction and its symmetry equivalents [23,24,25]. Typically all eight possible antiferromagnetic domains will form within the sample as it is cooled below $T_N$ (see fig 2a) so while longitudinal and transverse magnetic excitations can be defined within each domain they cannot be disentangled for the sample as a whole. However, when a magnetic field **B**, is applied along the crystallographic **b** direction of $KCuF_3$ a reorientation of spin moments occurs resulting in a spin-flop phase for fields above 0.8T [25,31]. In this phase the spin moments flop perpendicular to the **b** direction and due to the *x-y* anisotropy [22] they point along the **a** axis [31] resulting in a unique domain throughout the sample (see fig 2b). For a sample orientation where the **a** and **c** axis form the horizontal scattering plane of the instrument a vertical field magnet is used, longitudinal magnetic excitations will occur along the **a** axis, while transervse excitations will occur along the **b** and **c** directions. If measurements take place around (0, 0, 1.5), rule 4 implies that spin components in the **c** direction cannot be measured at all since they are parallel to the wavevector transfer and only the transverse component oscillating along the **b** direction is observable. For vertically polarized neutrons (i.e. neutron polarization parallel to **b**), longitudinal excitations which oscillate parallel to the **a** axis will be observed as SF scattering (rule 2)), and transverse excitations which oscillate parallel to the **b** will be observed in the NSF channel (rule 1) (see fig 2c). Phonons will of course also be observed in the NSF channel (rule 3). This arrangement is advantageous in cleanly separating longitudinal scattering from both transverse magnetic scattering and phonons and allowing the unambiguous determination of the nature of the signal. It should be noted that the field applied here (**B**=1T) is minimal compared to the exchange couplings and besides reorientation of the spin moments causes no further effects on the spectrum.

The polarized neutron measurements were performed on HB1. A flat monochromator crystal of FeSi - (1, 1, 0) reflection – was utilized, in conjunction with a flat Heusler analyzer crystal - (1, 1, 1) reflection. The neutron final energy was fixed at $E_F$=30.5 meV and a spin-flipper placed between sample and analyser allowed measurement of both NSF and SF channels to be made. Soller collimators with values 48´-120´-80´-240´ were used. In this configuration the resolution-limited spin-waves have FWHM in a constant-energy scan at 16 meV of 0.083Å$^{-1}$ (0.053r.l.u) and 0.056Å$^{-1}$ (0.036r.l.u) for the negative and positive branches respectively while the energy resolution for a non-dispersive mode at this energy is 3.2 meV FWHM. An Oxford Instruments vertical field, split coil, cryomagnet was used to generate a field of *B*=1T at the sample position. Measurements were made for sample temperatures of *T*=6 K and 200 K with temperature control to within +/-0.1 K. Some preliminary results of these measurements have been discussed elsewhere [12].

**V Results and Analysis**

*Unpolarized data*

Using the unpolarized setup a number of constant-energy and constant-wavevector scans were performed over the energy range 7 to 27 meV covering wavevectors from (0, 0, 1.3) to (0, 0, 1.7) to map out the magnetic scattering in the vicinity of the predicted longitudinal mode. Most measurements took place at low temperatures ($T<<T_N$) in the ordered phase where the longitudinal mode is predicted to exist, however the temperature dependence was also measured up to 70 K and scans were repeated at 200-300 K to obtain a background. Some of the unpolarised data was discussed in our previous paper (ref. [11]) however we include it also in this paper for clarity and completeness. Figure 3a shows a constant wavevector scan at the antiferromagetic zone centre, (0,0,1.5). The filled circles give the data measured well below $T_N$ and show a number of features including scattering close to the predicted longitudinal mode energy at 16 meV. When the measurement is repeated at *T*=200 K where the magnetic signal at low energies is significantly reduced by thermal broadening (open circles) the peaks at 19.5 meV and 25 meV remain identifying them as phonons. The scan at 200 K was used to determine the phonon contribution at low temperatures by dividing the phonon signal by the phonon thermal population factor at 200 K and multiplying by the population factor at 10 K, it was then smoothed and subtracted from the low temperature data as shown in fig 3b. Two magnetic features dominate – a large signal at low energies and a broad peak centred at 16 meV. The low energy scattering comes from the spin-waves which are increasingly captured by the resolution function at low energies as they disperse towards the antiferromagnetic zone centre. The peak lies close to the anticipated longitudinal mode energy although it is significantly broader than the predicted sharp resolution limited signal.

The peak was fitted to the theoretical prediction of the longitudinal mode (equation (5) above) [29], modified so that the linear dispersion parallel to the chain is replaced by a sinusoidal dispersion with a minimum at the antiferromagnetic zone centre to model the discrete nature of the lattice not included in the field theory calculations:

$$E_L = \sqrt{\frac{\pi^2 J^2}{4}\sin^2(2\pi L) + M^2\left(3 - \frac{\gamma(\cos(2\pi H) + \cos(2\pi K))}{2}\right)} \qquad (7)$$

The intensity was assumed to have a $1/E$ variation, as obtained by transforming the delta function in equation (6). The mode was convolved with the full four dimensional instrumental resolution and the observed scattering was characterised best using a Gaussian lineshape for the energy profile and fitted peak position of 14.9±0.1 meV. This energy gap is similar to the theoretical value of 17.4 meV [29] but quite different from the maximum of the two-magnon continuum predicted at 23.5 meV by spin-wave theory or from the lower edge of the two-spinon continuum predicted at 22 meV [29]. The mode is intrinsically broadened with FWHM of 4.95±0.4 meV, suggesting that its lifetime is shortened by decaying into spin-waves again making it different from both the two-magnon feature which would have a FWHM of 24 meV and the two-spinon continuum edge which would have an asymmetric lineshape.

It is important to eliminate the possibility that the feature at 16 meV is spin-wave signal that has been distorted by the resolution function to give the appearance of a mode. Figure 3c shows a constant energy scan at 16 meV through the antiferromagnetic zone centre (0, 0, -1.5). It shows two peaks due to the spin-wave branches which are fitted to the transverse mode dispersion again modified to take account of the sinusoidal dispersion.

$$E_T = \sqrt{\frac{\pi^2 J^2}{4}\sin^2(2\pi L) + M^2\left(1 - \frac{\cos(2\pi H) + \cos(2\pi K)}{2}\right)} \qquad (8)$$

The dashed line is a fit of equation (8) convolved with the instrumental resolution and including the $1/E$ intensity variation (equation (4)), where the only fitted parameter is the amplitude. The resolution function is accurately known for this spectrometer and it is clear that spin-wave theory does not account for all the scattering that occurs between the spin-wave peaks. Therefore the extra scattering must have some other origin.

A further indication that the additional scattering arises from the longitudinal mode comes from its temperature dependence. The longitudinal mode is associated with long-range magnetic order in the low temperature phase of a quasi-one-dimensional, spin-1/2, Heisenberg, antiferromagnet and should be absent above the Néel temperature where one-dimensional physics dominates and long-range order is lost. The intensity of the (0,0,1.5) magnetic Bragg peak is plotted as a function of temperature in fig 4a, this quantity is proportional to the square of the ordered moment and confirms the Néel temperature as 40K+/-1K. Fig 4b shows inelastic scans for a constant-wavevector of (0,0,1.5) at several temperatures, where the contribution from the phonon at 19.5 meV has been subtracted off. At $T$=6.15 K (filled circles), which essentially gives the ground state signal, the longitudinal mode is observed at 16 meV and the line through the data is a fit to spin-waves plus broadened longitudinal mode convolved with the spectrometer resolution. When the scan is repeated close to the Néel temperature at 40.6 K (open squares) the longitudinal mode is no longer a distinct entity and cannot be distinguish from the spin-wave scattering. Well above $T_N$ at 70.5 K (filled diamonds), the longitudinal mode has vanished and the observed spectrum shows the smooth decrease in intensity with increasing energy typical of the two-spinon continuum. The line through the data is the field theory expression for an ideal one-dimensional, spin-1/2, Heisenberg antiferromagnet at 70 K that has been corrected for interchain effects by a random phase approximation (equations (13), (17), (25) and (29) from ref. [28]) and convolved with the instrumental resolution; the only fitted parameters are the interchain exchange constant and overall amplitude.

It is clear from fig. 4b that at energies just below the longitudinal mode (8-11 meV) the signal shows a minimum at low temperatures but 'fills in' as temperature is increased until the distinction between the longitudinal mode and spin-wave scattering is lost. To investigate this further a constant-energy scan was done at 10 meV (fig 4c). At $T$=6K two spin-wave peaks are observed (filled circles) and the line through the data is a fit of the spin-wave branches convolved with the resolution function. Close to the Néel temperature at $T$=40.6K (open squares) the peaks have broadened and the region between them has partially filled in. Then at $T$=70.5K (filled diamonds) the two peaks are replaced by a single broad feature more characteristic of the two-spinon continuum than of spin-wave branches (again the line through the data corresponds to field theory plus random phase approximation [28]) and it is clear that the physics of a one-dimensional spin-1/2, Heisenberg antiferromagnet dominates at this temperature.

To make a direct comparison between the predicted and measured longitudinal mode, the full low temperature data set at $T$=10K was combined to show the neutron intensities as a function of wavevector close to $Q$=(0,0,-1.5) and with energy between 8 and 22 meV. The results are given in fig 5a where the colors represent neutron intensities. The spin-waves form the blue V-shaped rods dispersing from the zone centre and the longitudinal mode is the red band lying between the spin-wave branches at 16 meV. The longitudinal mode is however partially obscured by the phonon which lies at 19 meV and disperses downward away from the antiferromagnetic zone centre. The phonon extends across the figure both in-between and on either side of the spin-wave branches and has greater intensity on the higher wavevector side. The phonon can be modelled using high temperature data collected at 200K and corrected for the thermal population factors, the results are shown in fig 5b. Fig 5c gives the neutron data with the phonon subtracted off and shows clearly that much of the intensity in between the spin-waves around 16 meV remains. Figure 5d shows a simulation of the predicted magnetic scattering over the same energy and wavevector region. The longitudinal mode was assumed to follow the theoretical dispersion (equation (7)) and intensity (equation (6)) with a Gaussian profile of FWHM 4.95 meV and a zone centre energy gap fixed at 14.9 meV as obtained from fitting. The spin-wave branches were given a resolution limited profile. The lineshapes of the various modes were normalized so that the integrated intensity of the spin-waves was four times greater than that of the longitudinal mode as predicted theoretically [29]. The calculation includes corrections for the magnetic thermal population factor, the neutron scattering geometrical factor (see rule 4 of the experimental details section) and the $Cu^{2+}$ magnetic form factor. The similarity between figures 5c and 5d is striking and demonstrates not only the very real presence of the longitudinal mode but also the accuracy with which the theories predict its intensity relative to the transverse modes.

*Polarized data*

Whilst the unpolarized measurements clearly show signal at the expected longitudinal mode position and prove that it is real magnetic scattering rather than a phonon or a resolution effect, polarization analysis is required to demonstrate conclusively that it is due to fluctuations parallel to the ordered spin moment (longitudinal). The experimental setup described in section VI and figure 2 was utilized.

*(i) elastic measurements*

First the behaviour of magnetic Bragg peaks as a function of field was measured to determine the details of the spin-flop reorientation transition and to measure the instrumental flipping ratio. The intensity of a magnetic Bragg peak at $T$=0K is proportional to

$$S^{\delta\delta} \propto \sum_{\delta=a,b,c} \left(1 - \left(\frac{Q_\delta}{|\mathbf{Q}|}\right)^2\right) \times \sum_i \left|\langle G|S_i^\delta|G\rangle\right|^2 \qquad (9)$$

where $|G\rangle$ is the ground state wavefunction and $S_i^\delta$ is the component of the spin of the $i$th $Cu^{2+}$ ion in the direction $\delta$. As the applied magnetic field $B$ is increased the ordering direction of the spin moment changes from [1,1,0] (and equivalent directions) to [1,0,0] so for an arbitrary field strength the spin moments will lie between $a$ and $b$ axes making an angle $\theta$ with the $b$ axis, with $\theta$ varying from 45°($B$=0T) to 90° ($B$=0.8T). The components of the $i$th spin along the $a$, $b$ and $c$ axes are $\langle G|S_i|G\rangle \sin(\theta(B))$, $\langle G|S_i|G\rangle \cos(\theta(B))$ and 0 respectively; the components of the magnetic Bragg peak intensity in these directions are proportional to $\sin^2(\theta(B))$, $\cos^2(\theta(B))$ and 0. Due to the experimental setup described in the previous section the component along $a$ will be observed in the SF channel while the component along $b$ will be observed in the NSF channel (see fig 2c). Equation (9) also contains the geometrical factor $1-(Q_\delta/|\mathbf{Q}|)^2$, which states that only components of the magnetic order perpendicular to the wavevector transfer can be observed. The geometrical weighting for ordering along the $a$, $b$ and $c$ directions is given for four magnetic Bragg peaks in table 1.

| $\mathbf{Q}$ (r.l.u.) of magnetic Bragg peak | $1-(Q_a/|\mathbf{Q}|)^2$ | $1-(Q_b/|\mathbf{Q}|)^2$ | $1-(Q_c/|\mathbf{Q}|)^2$ |
|---|---|---|---|
| (0, 0, 3/2) | 1 | 1 | 0 |
| (-1, 0, 3/2) | 0.71 | 1 | 0.29 |
| (-2, 0, 1/2) | 0.07 | 1 | 0.94 |
| (-2, 0, 3/2) | 0.39 | 1 | 0.62 |

Table I  *Geometrical weighting, $1-(Q_\delta/|\mathbf{Q}|)^2$, of four magnetic Bragg peaks for the spin components measured in **a**, **b** and **c** directions.*

Combining the geometrical weightings with the components of magnetic order along the $a$, $b$ and $c$ directions gives the predicted intensities as a function of $\theta$ in the SF and NSF channels as shown in Table 2

| Magnetic Bragg peaks | $I_{NSF}(\theta(B))$ , $I_{SF}(\theta(B))$ | $I_{NSF}(45)$, $I_{SF}(45)$ [B=0.0T] |
|---|---|---|
| (0, 0, 3/2) | $\cos^2(\theta(B))$ , $1.00\sin^2(\theta(B))$ | 1/2 , 1/2 |
| (-1, 0, 3/2) | $\cos^2(\theta(B))$ , $0.71\sin^2(\theta(B))$ | 1/2 , 0.71/2 |
| (-2, 0, 1/2) | $\cos^2(\theta(B))$ , $0.07\sin^2(\theta(B))$ | 1/2 , 0.07/2 |
| (-2, 0, 3/2) | $\cos^2(\theta(B))$ , $0.39\sin^2(\theta(B))$ | 1/2 , 0.39/2 |

Table 2   *Predicted intensities for the four magnetic Bragg peaks in the SF and NSF channels as a function of field and normalized by $S^2$.*

The variation of the spin-flop angle *θ* with field *B* applied along the **b** axis is unknown except that for zero field the spin direction is [1,1,0] giving *θ*=45. The four magnetic peaks were measured as a function of field in 0.1T steps from *B*=0.0T to *B*=0.8T in the SF channel, and from this data *θ(B)* was deduced along with the efficiency of the neutron polarization or flipping ratio *FR(B)* as a function of field. These two quantities are plotted in fig 6 along with the Bragg peak intensities corrected from the deduced flipping ratio. The data shows that spin-flop reorientation is complete at *B*=0.8T with the ordered spin moments pointing entirely along the **a** axis.

*(ii) inelastic measurements*

The aim of the next part of the experiment was to determine the polarization of the signal observed at 16 meV. The techniques described above were used to create a monodomain sample by applying a magnetic field of *B*=1T and polarized inelastic neutron scattering measurements were performed on it. As with the unpolarized investigation, these measurements took place around the (0,0,1.5) magnetic Bragg peak, however because of the reduced intensities available with the polarized setup, we concentrated on just two scans – a constant-energy scan at 16 meV and a constant-wavevector scan at (0,0,1.5). The measurement took place in the ordered phase at *T*=6 K and data in both SF (longitudinal magnetic) and NSF (transverse magnetic plus phonons) channels was collected. The scans were also repeated at *T*=200 K to obtain a high temperature background and an analyser turned background was also measured. The first stage in processing the data was to correct for flipping ratio. The flipping ratio, FR, can be obtained from the ratio of NSF to SF scattering of a nuclear Bragg peak and for this we chose to measure the (0,0,2) peak. The Bragg scattering can of course only be measured at zero energy transfer and in order to find the flipping ratio as a function of energy the incident and final energies were kept equal to one another and changed from 30.5 meV to 61.5 meV in steps of 1 meV.

The data adjusted for FR is given in figs 7 and 8. Although the data has not been corrected for background or phonons there is a clear difference between the SF and NSF channels. In the constant-wavevector scans at *T*=6K (filled circles), the SF channel (fig 7a) which measures longitudinal magnetic scattering has a broad peak centred at 16 meV. This signal is not present at 200K (open squares) confirming the results of the unpolarized measurement. The signal is also absent in the NSF channel (fig 7b) which measures transverse magnetic scattering and phonons but not longitudinal magnetism. From these results it is clear that the signal at 16 meV is entirely longitudinal scattering. In fact this longitudinal feature is already clearly visible in the raw data (without the FR correction) which is given in ref. [12]. The NSF data has some structure at high energies due to the phonons observed previously in the unpolarized measurement. At low energies both SF and NSF data sets show increasing scattering; this is much stronger in the case of the NSF channel at low temperatures where the transverse spin-wave modes are captured by the resolution function as they disperse towards the zone centre. However a smaller quantity of low-energy signal is found in the SF channel and also at 200K in the NSF channel. This signal probably has its origins in the elastic incoherent scattering that is observable due to the broad energy resolution of this experimental setup. The incoherent scattering is observed in both the SF and NSF channels [30] and in each case the high temperature signal can be used as a measure of the incoherent background. The constant-energy scans are given in figure 8. A single peak is observed in the SF channel (fig 8a) at $L$=1.5 again corresponding to the predictions for the longitudinal mode whereas two peaks are found in the NSF measurement (fig 8b) on either side of this position corresponding to the transverse spin-wave branches. All of these scans were repeated with the analyzer turned away from its optimum position by 15 degrees so that signal from the sample is not Bragg scattered at the analyzer or directed towards the detector; this gives a measure of the general instrumental background in the absence of the sample which is significant due to the long counting times required.

The next stage in the data analysis was to subtract off the analyser turned background. This was done separately for the SF and NSF data sets where in both cases the analyser turned data set was fitted to a straight line which was subtracted from the data. This correction is successful in that it removes the slope from the constant-energy data and in addition accounts for almost the entire background of the SF measurement for energy transfers above 15 meV. Next the low energy incoherent signal and the phonons were subtracted. For the SF data the background was assumed to consist of incoherent signal only. which was obtained by fitting the constant-wavevector SF scan at 200 K and energies below 10 meV (where magnetic continuum signal is washed out due to temperature broadening) to a Gaussian function centred at zero energy transfer. This Gaussian is extrapolated over the range of the entire scan and is then subtracted from the low temperature SF data to obtain the longitudinal magnetic scattering. The background for the NSF data is more complex because, besides the incoherent component, the signal is also contaminated by phonons. As with the SF data the incoherent component is obtained by fitting the low energy data (below 10 meV) at 200 K in the constant-wavevector scan. The remaining scattering at 200 K in then assumed to be phonon signal plus a residual magnetic continuum contribution, but of course the relative proportions are unknown. We have made the assumption that the residual continuum is small and therefore that the remaining signal comes mostly from phonons - this will result in a small overestimation of the background. The phonon background at 6 K is worked out by adjusting the phonon signal at 200 K for the thermal population factors, it is then added to the incoherent background and the result is subtracted from the low temperature NSF data.

The results are shown in fig 9 where the analyzer turned, incoherent and phonon backgrounds have been subtracted so that the signal is entirely longitudinal magnetic in the case of the SF data and entirely transverse magnetic in the case of the NSF data. Fig 9a shows the constant-wavevector SF data at $T$=6K where the longitudinal mode is clearly visible and appears somewhat broader than in the unpolarized measurement due to the coarser energy resolution of the present setup. The solid line gives the fit of the longitudinal mode convolved with the resolution function, the energy gap extracted from the fitting is 14.55±0.24 meV while the FWHM of the mode is 6.0±0.6 meV. The complete absence of a peak at 16 meV in the NSF constant-wavevector scan at $T$=6K (fig 9b) confirms the longitudinal nature of this mode. The low energy signal in the NSF scan comes from the transverse spin-waves and has been fitted to a model of the spin-wave dispersion convolved with the instrumental resolution (solid line). Figure 9c shows the constant-energy scans in both the SF and NSF channels superimposed, again the peak at $L=1.5$ observed in the SF channel is completely absent in the NSF channel. From the fitting, the ratio of the intensity of the longitudinal mode to the spin-waves is 0.36±0.12 (integrated intensities are compared to take into account the broadened line width of the longitudinal mode) whereas from theory [29] this ratio is predicted to be 0.25. The larger ratio found experimentally may be due to an over-subtraction of the background for the NSF data which makes the spin-wave intensity appear smaller than it actually is.

While the longitudinal mode provides a good fit to the data in fig. 9a it is unable to account for the extra scattering that occurs above 23 meV. One explanation is that this signal comes from the multi-particle continuum which is predicted to have a lower boundary of 22 meV for $T<<T_N$ in $KCuF_3$ [29]. There is no exact expression for this continuum however we model it with a formula similar to the Müller ansatz [26] of the ideal one-dimensional Heisenberg antiferromagnet but with the signal truncated at the expected lower edge $E_c$ given by

$$E_c = \sqrt{\Delta_C^2 + \left(\frac{\pi J}{2}\right)^2 \sin^2(2\pi L)} \tag{10}$$

where $\Delta_C$ is the lower bound of the continuum at the antiferromagnetic zone centre and is predicted to have the value $2M$. The continuum scattering is then given by

$$\operatorname{Im}\chi_c = A_\delta \frac{1-\cos(2\pi L)}{2} \frac{1}{\sqrt{E^2 - \left(\frac{\pi J}{2}\right)^2 \sin^2(2\pi L)}} \Theta(E - E_C(L)) \tag{11}$$

where $\chi_c$ is the generalized spin susceptibility for the continuum, $\Theta(x)$ is the heaviside step function, $\delta$ is the polarization direction and $A_\delta$ is a constant of proportionality that is allowed to vary between transverse and longitudinal channels although theoretically it is independent of polarization. Fitting this expression along with the longitudinal mode to the SF data gives better agreement with the data. The fitted continuum signal is given by the dashed line figure 9a. The extracted value of $\Delta_C$ is 22.9±0.59 meV which is very similar to the predicted value of 22 meV implying that this extra scattering may well arise from a longitudinal continuum. Theory also predicts that the transverse continuum would occur at the same energy, this should be observed in the NSF channel. Looking at fig. 9b, it is not clear from the data that such a continuum exists. Introducing an additional transverse continuum to the NSF fitting function produces a small but insignificant improvement in the fit. We must however take into account that the NSF data is less reliable than the SF data, due to the phonon contamination and the greater uncertainties in the background subtraction and it is probably not possible to determine from the data whether or not a transverse continuum is present. The continuum in the low temperature antiferromagnetically ordered phase of $KCuF_3$ has been previously observed [5,10] and an estimate of its lower edge, given by the onset of the energy/temperature scaling predicted for the magnetic correlations of an ideal ID spin-1/2, Heisenberg antiferromagnet [32], was found to be 26 meV [10]. There are similarities between the low temperature phase of $KCuF_3$ and the spin-Peierls phase in $CuGeO_3$ where a gapped dimer mode is observed, separated by a further gap from a continuum [33]. In each case the perturbation that drives the system from ideal 1D behaviour (weak staggered magnetic order in $KCuF_3$ and alternating intrachain exchanges in $CuGeO_3$) forces the continuum to be gapped while an additional mode appears within that gap.

Finally, while theoretically the longitudinal mode is predicted to be a distinct entity from the longitudinal continuum an alternative view point is that it is simply the lower bound of the continuum. We have explored this issue by fitting only the continuum function to the data using a lower bound energy of 13 meV (dotted line fig 9a). While the continuum can accurately model the lower energy side of the longitudinal mode, it predicts a decrease in intensity that is far too gradual at higher energies so that the simulation lies well above the data points for energies above 18 meV. It is therefore clear that the conventional continuum function is unable to account for the scattering at 16 meV and that we should think of this scattering as a broadened mode rather than the continuum edge.

**VI Discussion and conclusion**

The results show both polarized and unpolarized neutron scattering data of the low energy excitation spectrum of the quasi-one-dimensional, $S=1/2$, Heisenberg antiferromagnet $KCuF_3$ in its long-range magnetically ordered state. The data reveal transverse spin-wave excitations and an additional mode. By using neutron polarization analysis to separate longitudinal and transverse magnetic excitations, we are able to demonstrate the longitudinal nature of this mode unambiguously. The spin-waves are resolution limited and disperse steeply downwards toward the antiferromagnetic zone centre, while the longitudinal mode is gapped with a gap size of ~15 meV that is fairly similar to the theoretically predicted value of 17.4 meV [29]. One feature of the longitudinal mode not predicted by theory is its sizeable broadening of approximately 5-6 meV suggesting that it is unstable to decay into spin-waves. However the intensity of the mode relative to the spin-waves is in good agreement with theory, the experimental ratio is 0.36±0.12 compared to the theoretical ratio of 0.25. Theory also predicts a two-spinon continuum at higher energies with a lower boundary of 22 meV, and data collected in the spin-flip (longitudinal) channel does indeed suggest that a longitudinal continuum starts at 23 meV. However in the non-spin-flip (transverse) channel, uncertainties in the background subtraction make it difficult to determine whether or not the expected onset of the transverse continuum is present at a similar energy.

It is interesting to compare these results to those for the longitudinal excitations in the compound $BaCu_2Si_2O_7$. This material is also a quasi-one-dimensional, $S=1/2$, Heisenberg antiferromagnet however its inter-chain interactions are much weaker. As a consequence the ordered spin moment per $Cu^{2+}$ ion for $BaCu_2Si_2O_7$ is $0.30\mu_B$ compared to $0.54\mu_B$ for $KCuF_3$ and its Néel temperature expressed in terms of the intrachain exchange constant is $0.033J/k_B$ compared to $0.099J/k_B$ [27,34]. The excitations in $BaCu_2Si_2O_7$ have been investigated using unpolarized neutron scattering and the longitudinal magnetic intensity has been inferred by comparing data above and below a spin-flop transition that causes the direction of the ordered spin moment to change. The evidence for a longitudinal mode is less conclusive for in this material. As for $KCuF_3$, longitudinal scattering is observed at the antiferromagnetic zone centre for energies well below the expected longitudinal continuum lower boundary, however the signal is very broad in energy. When a function consisting of the longitudinal mode plus longitudinal continuum is fitted to the data the longitudinal mode has an energy gap of 2.1 meV and width of 1.5 meV, while the lower boundary of the longitudinal continuum is 5 meV [27,34]. For this compound the width of the longitudinal mode compared to its energy gap is 0.71 and thus the relative broadening is much greater than for $KCuF_3$ where the ratio is 0.34, resulting in considerably more overlap between the longitudinal mode and the continuum for $BaCu_2Si_2O_7$ than $KCuF_3$. Furthermore the intensity of the longitudinal mode in $BaCu_2Si_2O_7$ normalized to the spin-wave intensity is four times greater than that predicted theoretically [27,34] in contrast to $KCuF_3$ where theory and experiment agree within error. In fact the longitudinal excitations in $BaCu_2Si_2O_7$ can be fitted almost as well by a model where there is no longitudinal mode but only a longitudinal continuum whose lower boundary is 2.0 meV [27,34]. Again this is in sharp contrast to $KCuF_3$ where a continuum only explain completely fails to model the data (see figure 9a).

There are several possible reasons for the discrepancy in the results for these two compounds. To begin with, $BaCu_2Si_2O_7$ has a more complex Hamiltonian (consisting of two zigzag magnetic chains and three interchain interactions) giving rise to a much more complicated structure factor in the excitation spectrum. It should be noted that any lifetime broadening of the longitudinal mode via decay into spin waves depends crucially on details of the interchain couplings. For chains coupled in a two-dimensional plane no longitudinal mode exists as the phase space for decay is too large; therefore the significant deviation of $BaCu_2Si_2O_7$ from the canonical three-dimensional tetragonal Hamiltonian, (equation (2)), may be important to the lifetime (and width) of the mode.

Another factor is that the definitude of the longitudinal mode concept varies with the degree of one-dimensionality of the material. In the strictly one-dimensional limit there is no longitudinal mode. Conversely, in the fully three-dimensional limit the mode is pushed to arbitrarily high energy and its contribution to the dynamic susceptibility is negligible. The posited longitudinal mode is only measurable in practice in systems with an intermediate inter-chain coupling strength. A sufficient amount of interchain coupling is presumably necessary to separate the mode from the spin-waves and to stabilize it. Too much interchain coupling however increases ordered spin moment and thus concentrates the longitudinal spectral weight in the magnetic Bragg peaks rather than the longitudinal mode. In addition the width of the mode would also increase as more of the spin-wave states lie below it in energy and become available for it to decay into. The precise behaviour of the dynamical correlations as a function of interchain coupling is unknown. For $KCuF_3$ the coupled chain RPA theory accurately predicts the values for the intensity and energy gap but does not account for the broadened linewidth of the mode produced by decaying into spin-waves. The interchain interactions in $BaCu_2Si_2O_7$ may well be sufficiently weak that the apparent physical behaviour is in a different regime where the mode, if present, is much more difficult to observe or to disentangle from the continuum scattering.

In conclusion our work combined with that on $BaCu_2Si_2O_7$ shows that longitudinal excitations occur below the predicted continuum lower boundary in a quasi-one-dimensional, spin-1/2, Heisenberg antiferromagnet. It is clear that in $KCuF_3$ these excitations can be thought of as a longitudinal mode with a broadened linewidth. The range of interchain to intrachain magnetic couplings for which a well-defined mode should be observable remains an open question, and measurements on other materials will be needed to fully delineate the possible physical regimes.

**Acknowledgements**
We thank F.H.L. Essler, A.M. Tsvelik and A. Zheludev for helpful discussions, and G. Shirane for the loan of the crystal. ORNL is operated by UT-Battelle LLC., under contract no. DE-AC05-00OR22725 with the U.S. Department of Energy.

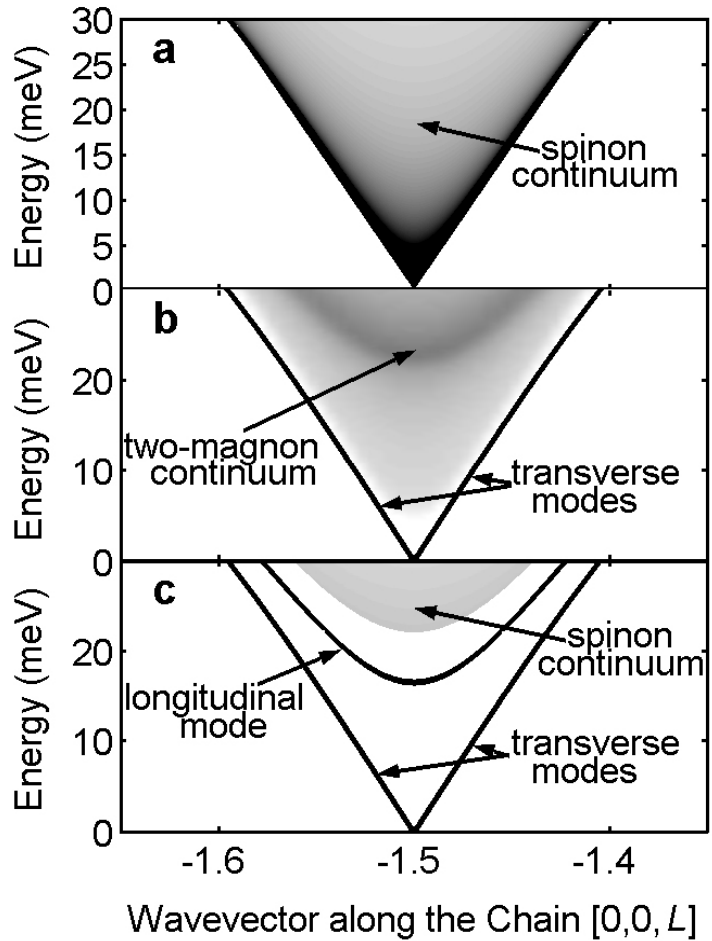

**Figure 1**

Predicted $T=0$ low energy magnetic response near the $(0,0,L)$ 1D antiferromagnetic wavevector in $KCuF_3$. The darkness of the shading gives the intensity of the scattering. The theories illusrated are a) The Muller ansatz. b) Spin-wave theory up to second order, showing one-magnon signal (black lines) and two-magnon scattering (shading). c) Predictions of Essler *et al* [29] and Schulz [18] showing transverse spin-waves and the longitudinal mode at low energies, and continuum at higher energies.

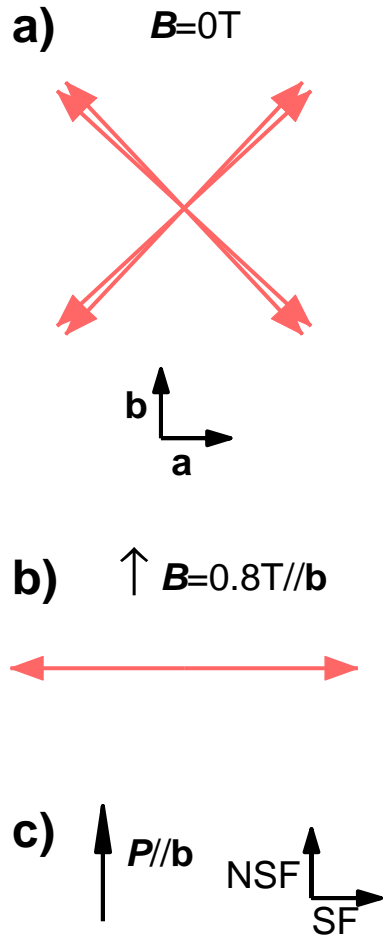

**Figure 2**

The magnetic ordering in $KCuF_3$ and the experimental setup. Below the Néel temperature in zero field the spin moments of $KCuF_3$ lie in the **a-b** plane and are canted slightly away from the [1,1,0] direction giving a total of eight possible antiferromagnetic domains (see panel a)). If however a magnetic field is applied along the **b** direction the spin moments lie parallel to the **a** axis for fields greater than 0.8T (see panel b)). In the experiment the neutrons are polarised parallel to the **b** axis. Components of the magnetic ordering that are parallel to the polarisation (***P***) direction are seen in the NSF channel while components perpendicular to ***P*** are observed in the SF channel (see panel c)). In zero field the multidomain structure means that magnetic Bragg peak intensity is observed in both the NSF and SF channels, however the single domain achieved with ***B***=0.8T//**b** means that this signal is observed solely in the SF channel. Longitudinal magnetic excitations (parallel to ordering direction) can also be separated from transverse magnetic excitations (perpendicular to ordering direction) using this technique.

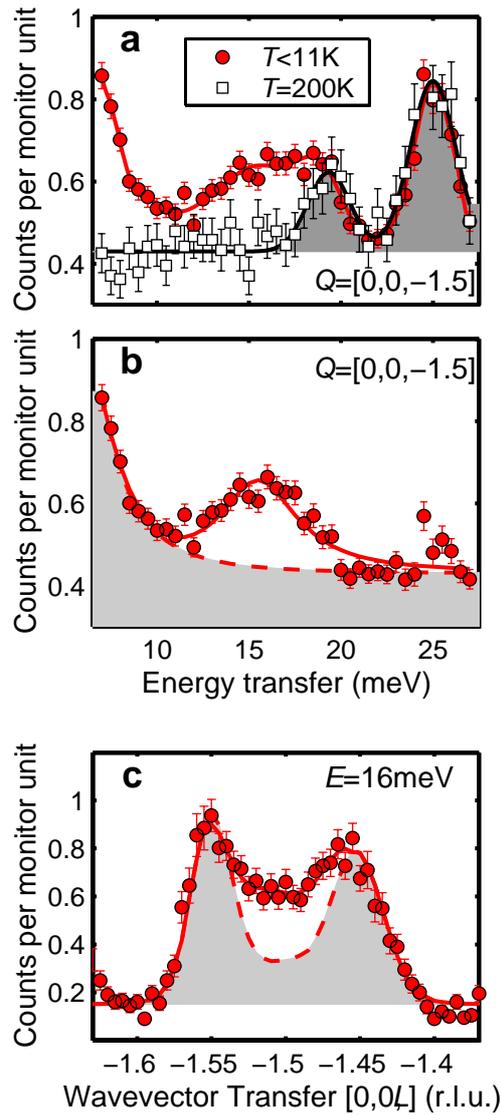

**Figure 3**

Constant-energy and constant-wavevector scans through the predicted longitudinal mode position using unpolarized neutron scattering. a) Constant-wavevector scan through (0, 0, -1.5) below $T_N$ at $T<11$ K (circles) and at $T=200$ K (open squares), the data is given per unit monitor (~1s). The high temperature data is used to identify the phonons (grey shading) and the solid lines are guides to the eye. b) Magnetic signal for $T<11$ K where the phonon background (adjusted for thermal populations factors) has been subtracted from the low temperature data. The scan consists of transverse magnetic signal (grey shading) and a lump of scattering at 16 meV close to the expected longitudinal mode position. c) Constant-energy scan at 16 meV, again the grey shading gives the scattering expected from the transverse spin-waves. In parts b) and c) the dashed line is the spin-wave intensity while the solid line is the fitted intensity of both the spin-waves and the longitudinal mode convolved with the instrumental resolution.

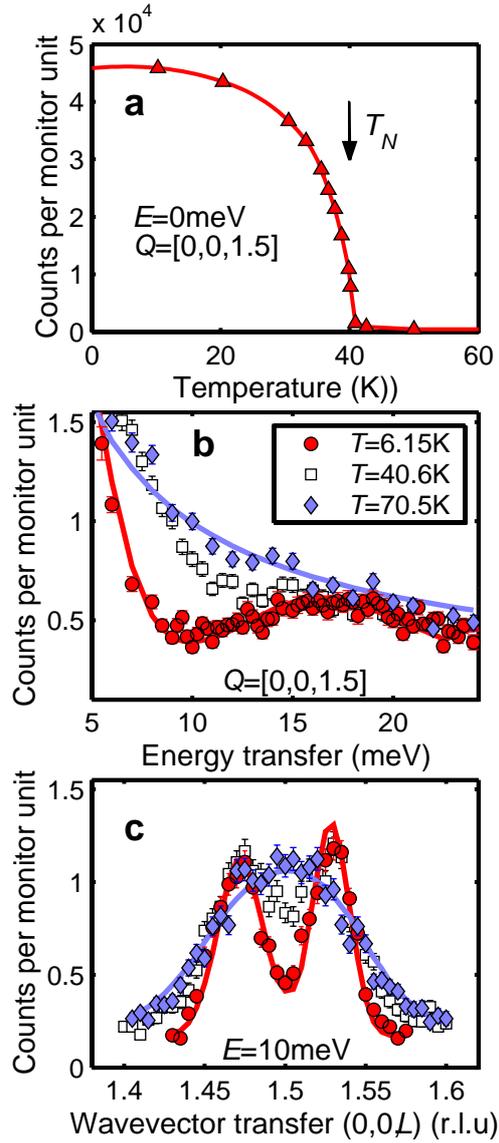

**Figure 4**

Temperature-dependence of the magnetic order and excitations measured with unpolarised neutrons. a) Intensity of the (0,0,1.5) magnetic Bragg peak as the crystal is heated through its Néel temperature $T_N$=39 K. The line through the data is a guide to the eye. b) Magnetic excitations at (0,0,1.5) for the temperatures 6.15 K (filled circles), 40.6 K (open squares) and 70.5 K (filled diamonds). c) Constant-energy scan below the longitudinal mode energy at 10 meV. In parts b) and c) the lines through the $T$=6.15 K data are fits of the transverse spin-waves and broadened longitudinal mode [26] convolved with the spectrometer resolution. The lines through the $T$=70.5 K data are fits of the field theory expression for an ideal, one-dimensional, spin-1/2, Heisenberg antiferromagnet at 70 K that has been corrected for interchain effects by a random phase approximation [28] and convolved with the resolution.

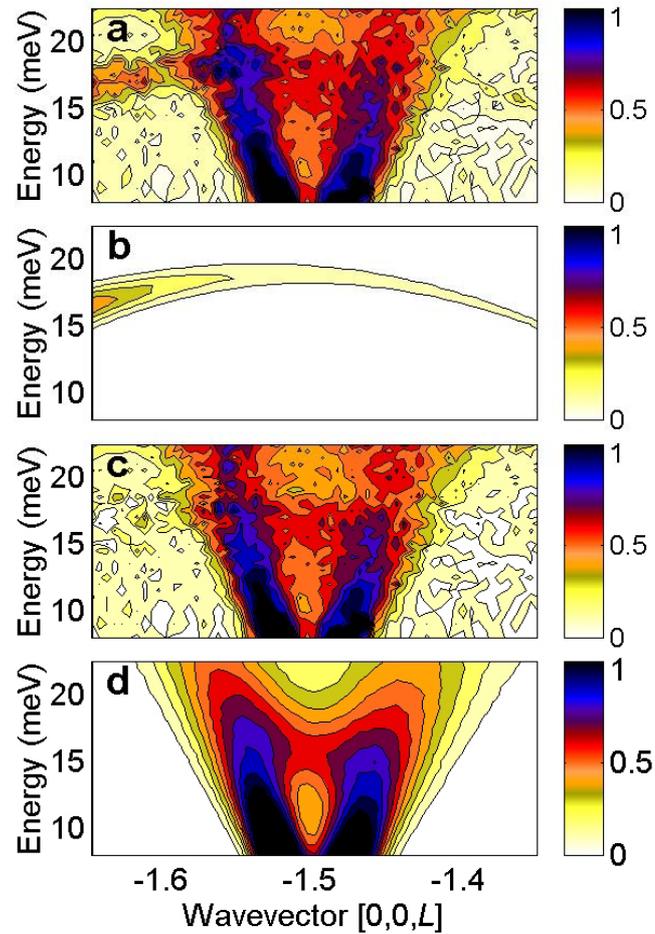

**Figure 5 (color)**

Wavevector–energy intensity maps over the longitudinal mode region measured using unpolarised neutrons. Relative scattering intensity is indicated by color as defined in the color bars. a) Raw data at $T<11$K. The most intense scattering is in the V-shaped spin-wave branches dispersing from the antiferromagnetic zone centre. The longitudinal mode at 16 meV is readily visible near (0,0,-1.5), but the data is contaminated by the 19 meV phonon. b) The phonon scattering modelled using the high temperature data corrected for the thermal population factor. c) Low temperature data with the phonon scattering subtracted off. d) Theoretical scattering intensity.

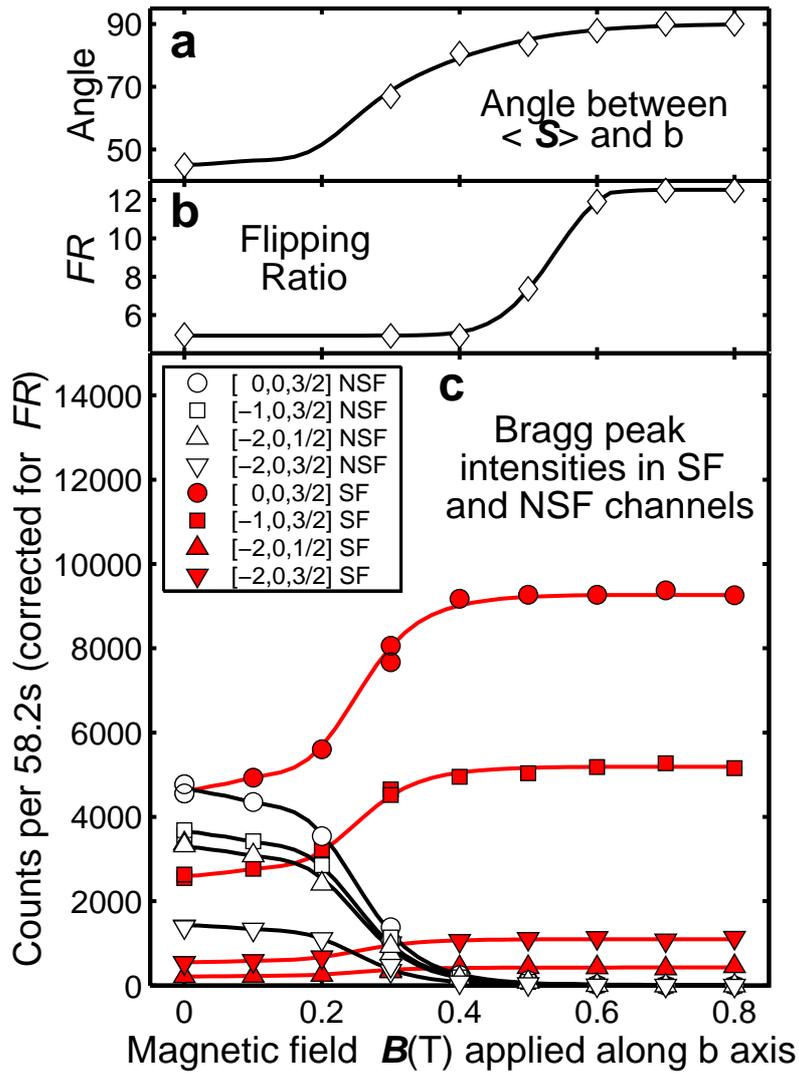

**Figure 6**

Magnetic Bragg peak intensity measured with polarised neutrons. A magnetic field applied along the **b** axis is used to pole the crystal into a single magnetic domain where all the spins point along the **a** axis. The intensities of the magnetic Bragg peaks vary as the spin direction changes. The intensities of four magnetic Bragg peaks are plotted in part c) for both SF and NSF channels. Saturation is achieved for fields greater than 0.8T. The angle $\theta$ that the spin moments make with the **b** axis is given in a) and the flipping ratio is also deduced and is plotted in b).

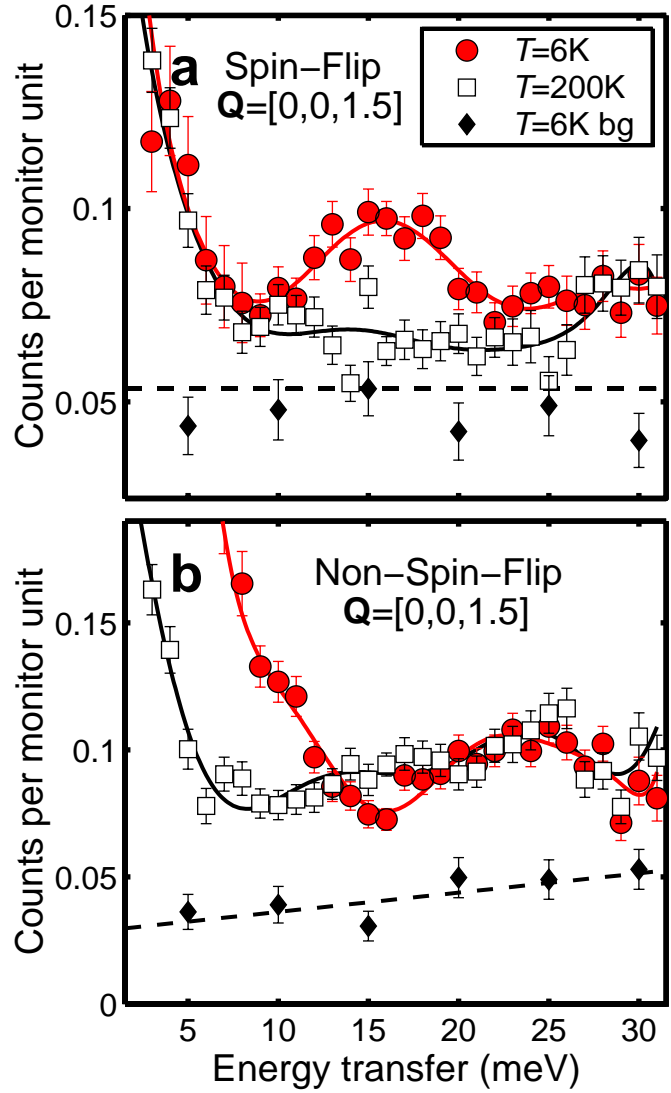

**Figure 7**

Constant-wavevector scans collected at $Q$=[0,0,1.5] using polarized neutron scattering data corrected for flipping ratio. The data was collected at $T$=6K (filled circles), $T$=200K (open squares) and $T$=6K with analyzer turned by 15 degrees from its optimum position (filled diamonds) to give a measure of the general instrumental background. Longitudinal magnetic scattering is observed in the spin-flip channel (part a)) while transverse magnetic scattering plus phonons are observed in the non-spin-flip channel (parts b)). The lines through the data are guides to the eye.

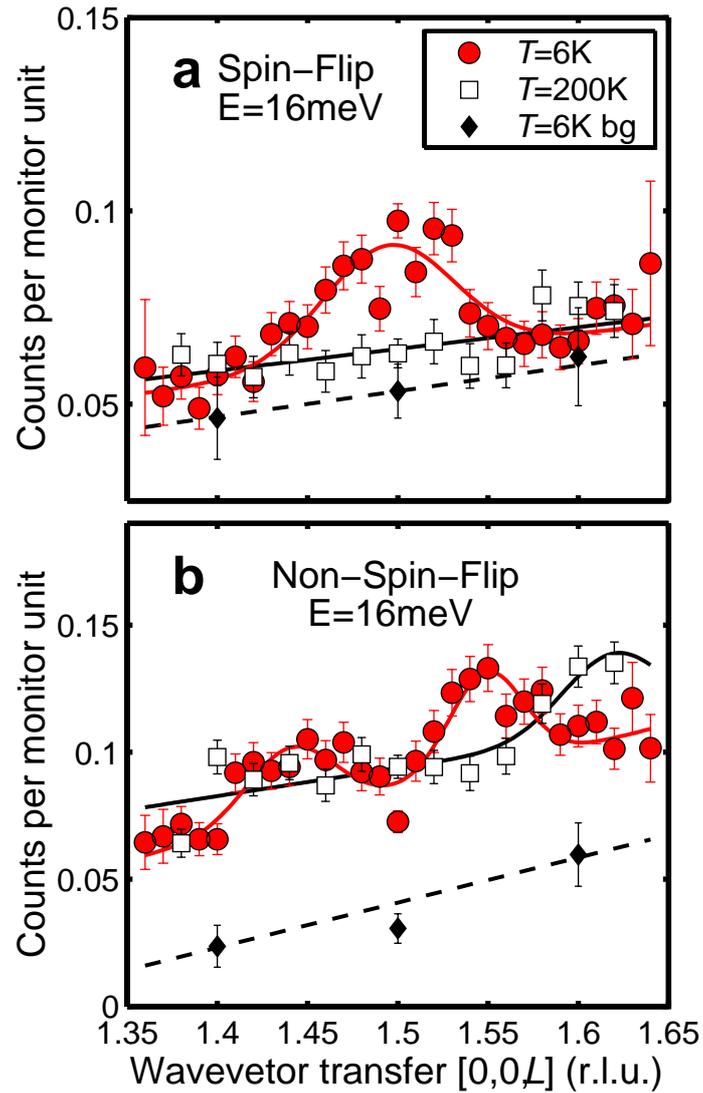

**Figure 8**

Constant-energy scans collected at *E*=16meV using polarized neutron scattering data corrected for flipping ratio. As in figure 6 the data was collected at *T*=6K (filled circles), *T*=200K (open squares) and *T*=6K with analyzer turned by 15 degrees (filled diamonds). Longitudinal magnetic scattering is observed in the spin-flip channel (part a)) while transverse magnetic scattering plus phonons are observed in the non-spin-flip channel (parts b)). Again the lines through the data are guides to the eye.

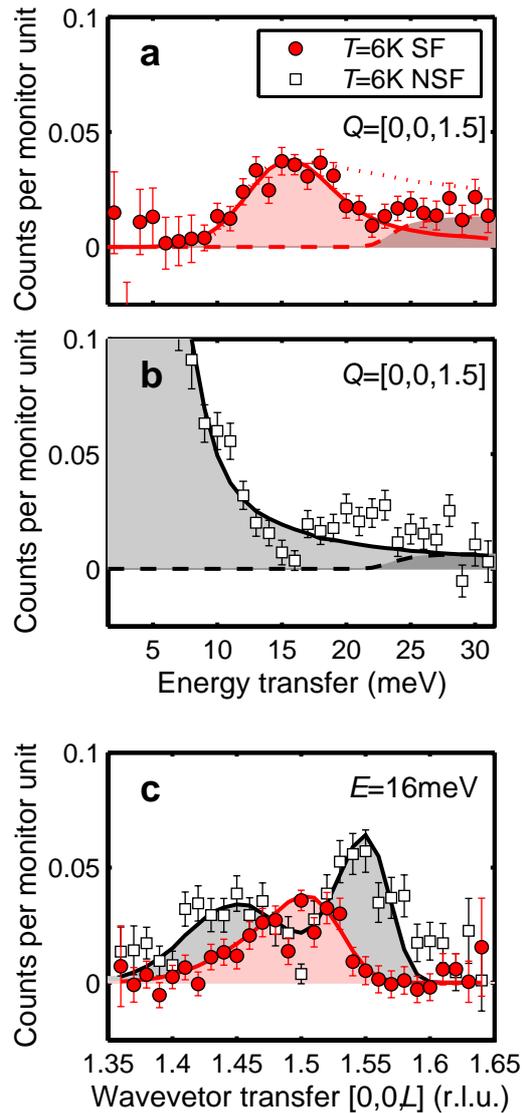

**Figure 9**

Polarized neutron scattering data corrected for flipping ratio and with the analyzer turned, incoherent and phonon backgrounds subtracted. a) Constant-wavevector scan collected at $Q$=[0,0,1.5] and $T$=6K and measured in the SF channel. The longitudinal mode and the longitudinal continuum are fitted simultaneously to the data; the solid line gives the longitudinal mode part of this fit while the dashed line gives the longitudinal continuum part. In an alternative model the scattering observed at 15 meV is due to the lower edge of the longitudinal continuum; the dotted line is a simulation of this. b) Constant-wavevector scan collected at $T$=6K in the NSF channel. The transverse spin-waves and transverse continuum have been fitted to the data and the solid line is the spin-wave part of this fit while the dashed line is the transverse continuum part. c). Constant-energy scans at $T$=6K in both SF and NSF channels. The solid lines through the data are the fitted longitudinal mode and spin-waves. The asymmetry in the spin-wave peaks on either side of [0,0,1.5] is a consequence of instrumental resolution. In all cases the resolution function has been convolved with the fitted lineshape.